\documentclass[twocolumn,showpacs,preprintnumbers,prb]{revtex4}
\usepackage{epsf}

\begin{document}

\preprint{ LPTENS-02-17}

\title{Critical Behavior and Lack of Self Averaging in the Dynamics of the
Random Potts Model in Two Dimensions}

\author{C.~Deroulers}
%\affiliation{
%Laboratoire de Physique Th\'eorique de l'\'Ecole Normale Sup\'erieure,
%24 rue Lhomond, 75231 Paris Cedex 05, France}
\affiliation{
Laboratoire de Physique Th\'eorique de l'\'Ecole Normale 
Sup\'erieure\footnote{Unit\'e Mixte de Recherche 8549 du Centre
National de la Recherche Scientifique et de l'\'Ecole Normale
Sup\'erieure}, 24 rue Lhomond, 75231 Paris Cedex 05, France}
\email{derouler@mozart.ucsc.edu}
\author{A.~P. Young}
\affiliation{Department of Physics, 
University of California, Santa Cruz CA 95064, USA}
\affiliation{Department of Theoretical Physics, 1 Keble Road,
Oxford OX1 3NP, England}
\email{peter@bartok.ucsc.edu}

\date{\today}

\begin{abstract}
We study the dynamics of the $q$-state random
bond Potts ferromagnet on the square lattice
at its critical point by
Monte Carlo simulations with single spin-flip dynamics.
We concentrate on
$q=3$ and $q=24$ and find, in both cases,
conventional, rather than activated, dynamics. We also look at
the {\em distribution}\/ of relaxation times among different samples, finding
different results for the two $q$ values. For $q=3$ the relative
variance of the relaxation time $\tau$ at the critical point
is finite. However, for $q=24$ this
appears to diverge in the thermodynamic limit and it is $\ln\tau$
which has a finite relative variance.  We speculate that this difference occurs because
the transition of the corresponding pure system is second order for $q=3$ but
first order for $q=24$. 

\end{abstract}

%\preprint  PPreprint: LPTENS-02-17
%\preprint: Preprint: LPTENS-02-17

\maketitle

\section{Introduction}
\label{sec:intro}
The
random $q$-state Potts model\cite{wu} in two-dimensions
has recently been the subject of extensive study. 
One reason for this interest
is that disorder even changes the
{\em order}\/ of the transition for $q > 4$, where it is first order in the
pure case but must be continuous\cite{imry-wortis,a-w,h-b}
% Small change
in two-dimensions for all-$q$. 
Another reason 
interest is the
prediction by Ludwig\cite{ludwig}, later
confirmed by numerical simulations\cite{cardy,jacobsen,terry},
of multi-fractal
exponents at the critical point.
Although there has been a lot of 
numerical work on the static critical
exponents\cite{cardy,jacobsen,terry,chatelain-prl,chatelain,cb99,cb01,picco,picco98},
there has been
less on the dynamics, though there have been some recent studies based
on the ``short-time dynamics''\cite{yh00,pan,ybjls} approach.
Here we perform a careful analysis of the dynamics of the Potts model in two
dimensions by Monte Carlo simulations. Our motivation is two-fold:

Firstly, Cardy\cite{cardy-paris} has
raised the possibility that, at least for certain types of dynamics,
one might have {\em activated}\/ dynamical scaling, in which
the log of the characteristic time $\tau$
(i.e. the barrier height) varies with a power of an
appropriate length scale $l$
i.e. $\ln \tau \sim l^\psi$ where $\psi$ is the ``barrier''
exponent. This is in contrast to
{\em conventional}\/ dynamical
scaling in which
$\tau$ itself varies with a power of $l$. i.e.
$\tau \sim l^z$, where $z$ is the dynamical exponent.
Activated dynamics has been proposed by Fisher\cite{fisher} for the
random field Ising model where the fixed point (of the renormalization group)
is at $T=0$. First order transitions can also be described by a $T=0$ fixed
point and also have activated dynamics.
In that case the barrier
exponent $\psi$ is equal to $d-1$ for a discrete broken symmetry,
where $d$ is the dimension, so $\psi=1$ here.

% Change
We study two values
of $q$: $q=3$ where the pure system has a second order transition and so 
activated dynamics seems unlikely for the random case, and $q=24$
where the
pure system has a first order transition so there is a greater
possibility that the random problem will have activated dynamics.
The size of the first order jump
increases as $q$ gets large, and we find that we need $q \ge 16$ to clearly
see activated dynamics for the {\em pure}\/ system for the range of sizes
studied.
% Small change
Hence, for the {\em random}\/ case, we take a rather large number of Potts
states, $q=24$, since activated dynamics is easily seen in this case for the
pure system, so we feel we have a good chance of seeing it in the random case
too, if it occurs. In fact, we find conventional dynamical scaling for both
values of $q$, but, as we shall discuss, have some concern as whether this is
the true asymptotic behavior for $q = 24$.

Secondly, there has recently been considerable discussion about lack of self
averaging for static quantities in random systems at the critical point. It is
found\cite{ah96,ahw98,wd98} that if disorder is irrelevant, then the relative
variance, $R_X$, of a static quantity $X$ tends to zero in the thermodynamic
limit like $L^{\alpha/\nu}$, where $L$ is the lattice size, and
$\alpha \ (< 0)$ and $\nu$ are the specific
heat and correlation length exponents, respectively. This is called
weak self averaging. However, if disorder is relevant then $R_X$
tends to a universal constant, indicating lack of self-averaging. In this paper
we consider the question of
self averaging for {\em dynamics}\/, which has not been discussed before,
to our knowledge. Disorder is relevant for both values of $q$ that we study,
and yet we find rather different distributions of the relaxation
time $\tau$ in the two
cases.
For $q=3$
we find a finite
relative variance for $\tau$
but, perhaps surprisingly, for $q=24$ 
this
appears to diverge in the thermodynamic limit and it is $\ln\tau$
which has a
finite relative variance.

Sec.~\ref{sec:model} describes the model and the quantities we calculate.
Results for the pure models are given in Sec.~\ref{sec:pure} while results for
the random system are given in Sec.~\ref{sec:random}. Our conclusions are
summarized in Sec.~\ref{sec:conclusions}.

\section{The Model}
\label{sec:model}
The Hamiltonian of the $q$-state Potts model is given by
\begin{equation}
\label{ham}
\beta {\cal H} = -\sum_{\langle i, j\rangle} K_{ij} \delta_{n_i n_j} ,
\end{equation}
where each site $i$ on an $N = L\times L$ square lattice is in one of
$q$-states, characterized by an integer $n_i =1,2,\cdots,q$. The couplings,
$K_{ij}$, are positive, and
include the factor of $\beta \equiv 1/k_B T$. They are independent
random variables, drawn from a probability distribution, $P(K)$.

For the pure case, the system is at criticality if the ``dual coupling'' $K^*$,
defined by\cite{wu,k-d}
\begin{equation}
( e^K - 1) ( e^{K^*} - 1) = q ,
\end{equation}
is equal to $K$, i.e. $K^* = K = K_c$, where
$e^{K_c} = 1 + \sqrt{q}$.
For the random case it is possible to choose a 
self-dual {\em distribution}\/ of the couplings to 
ensure that the system is at the critical point.
We take
the distribution suggested by Olson and Young\cite{terry} (OY) since this is
both very broad (so the system is far from the pure fixed point), and
also does
not have a substantial weight near $K=0$ (so the system is far from the
percolation fixed point). In terms of $x \equiv e^{-K}$, the OY distribution
is\cite{correction}
\begin{equation}
P_X(x) =  {2 \over \pi}\  { \sqrt{q} \over (1-x)^2 + q x^2 } .
\label{px}
\end{equation}
To generate random numbers with probability $P_X(x)$ 
one takes $x$ to be
\begin{equation}
x = {1 \over 1 + \sqrt{q} \tan (\pi r / 2) } ,
\end{equation}
where $r$ is a random number with a uniform distribution between 0 and 1. We
expect that the same results would be obtained asymptotically for {\em any}\/
reasonable distribution. However, to verify this would require a very large
computational effort, since other distributions are likely to have larger
corrections to scaling than the OY distribution, and so would need simulations
on larger lattices than we have been able to study here. We have some
preliminary results for a binary distribution which are similar to those
presented here for the OY distribution, but a much larger computational effort
would be needed to verify convincingly that the two distributions are 
indeed in the
same universality class. 

We focus on the time dependent magnetization squared, defined by
\begin{equation}
\label{m2-def}
m(t)^2 = {q \over q-1} \left( \sum_{n=1}^q \rho_n(t)^2 - {1\over q}
\right) , 
\end{equation}
where $\rho_n(t)$ is the fraction of sites in state $n$ at time $t$.
The initial configuration is completely random and so $m(t)^2$ increases,
eventually reaching its equilibrium value for $t \to \infty$.
Note that
$m(t)^2$ is invariant under global symmetry transformations which permute the
states $n$.
Particularly useful is the average value of $m(t)^2$
normalized by its equilibrium value $m(\infty)^2$,
i.e.
\begin{equation}
\widehat{m}(t)^2 = {[ \langle m(t)^2 \rangle ]_{\rm av} \over
[ \langle m(\infty)^2 \rangle ]_{\rm av} } ,
\label{m2}
\end{equation}
where $[\cdots]_{\rm av}$ denotes an average over disorder, and $\langle
\cdots\rangle$ denotes a thermal average. Clearly $\widehat{m}(0)^2 \simeq 0$ (for
finite-$N$ it is $O(1/N)$), and
$\lim_{t\to\infty} \widehat{m}(t)^2 = 1$.

In addition to results for the {\em average}\/ decay of the magnetization, we
also discuss the {\em distribution}\/ of relaxation times of the total
magnetization.

Since we want to use a realistic form for the dynamics, we use standard Monte
Carlo methods rather than one of the more efficient cluster
algorithms\cite{sw,wolff} that reduce critical slowing down. We expect that
any ``local'' dynamics would give the same results, but that cluster algorithms,
in which the average size of the cluster of flipped spins diverges at the
critical point, would yield faster dynamics, as indeed they are designed to
do.  We therefore used the Wolff\cite{wolff} cluster algorithm to obtain more
accurate estimates for some of the {\em equilibrium}\/ values of $m^2$, needed
in the denominator of Eq.~(\ref{m2}). 

\section{Results for the pure system}
\label{sec:pure}
\begin{figure}
\begin{center}
\epsfxsize=\columnwidth
\epsfbox{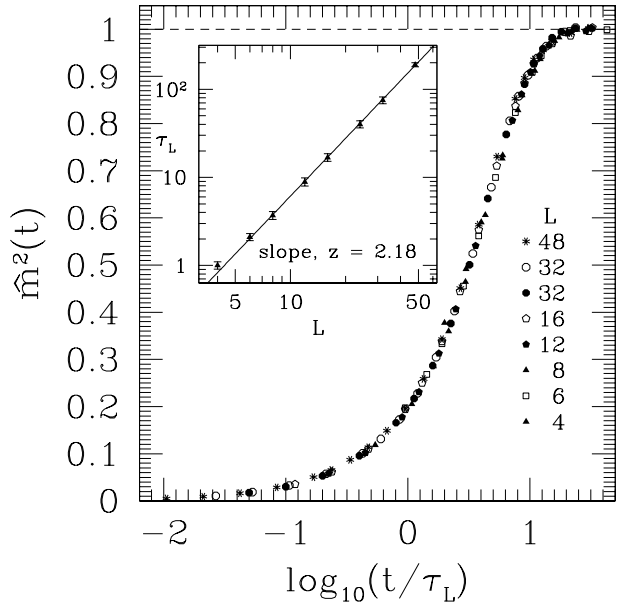} 
\end{center}
\caption{
Results for $\widehat{m}(t)^2$, defined in Eqs.~(\ref{m2}) and (\ref{m2-def}),
for the pure system for $q=3$. The horizontal axis
is $t/\tau_L$, where $\tau_L$
is determined for each lattice size by requiring that the data collapse.
%Results for the $\tau_L$ are shown in Fig.~\ref{pure_3_tau}.
The inset shows the resulting values for
$\tau_L$.
%determined so that the
%data in the main part of the figure collapses.
Since all the $\tau_L$ can be multiplied
by the same constant with no effect on the quality of the data collapse, we
arbitrarily set $\tau_4 = 1$. The best fit, omitting the $L=4$ point, gives 
$z = 2.18 \pm 0.04$.
}
\label{pure_3_plus_tau} 
\end{figure}

The transition is second order for $q <4$, so we expect conventional
dynamical scaling in this region. To see this we show in
Fig.~\ref{pure_3_plus_tau} data for 
$\widehat{m}(t)^2$ for $q=3$ against $\ln (t/\tau_L)$, where $\tau_L$ is chosen for each
$L$ in order to collapse the data. A log-log plot of $\tau_L$ against $L$
is shown in the inset to Fig.~\ref{pure_3_plus_tau},
which demonstrates nice power-law scaling. A fit, omitting the $L=4$ data
point, gives a
dynamical exponent $z =2.18 \pm 0.04$. An activated scaling plot of
$\ln \tau_L$ against $L^\psi$ does not work unless $\psi$ is
chosen to be extremely small,
in which case this is equivalent to a conventional scaling plot. 

%\begin{figure}
%\begin{center}
%\epsfxsize=\columnwidth
%\epsfbox{pure_3_tau.eps} 
%\end{center}
%\caption{
%Results for $\tau_L$ for the pure system for $q=3$, determined so that the
%data in Fig.~\ref{pure_3} collapse. Since all the $\tau_L$ can be multiplied
%by the same constant with no effect on the quality of the data collapse, we
%arbitrarily set $\tau_4 = 1$. The best fit, omitting the $L=4$ point, gives 
%$z = 2.18 \pm 0.04$.
%}
%\label{pure_3_tau} 
%\end{figure}

By contrast, for $q >4$ the transition is first
order which leads to activated scaling with $\psi = 1$, as discussed in
Sec.~\ref{sec:intro}.  The correlation length at the critical point has been
computed in Ref.~\onlinecite{buffenoir} and we show a table of results,
computed from the expressions in this reference, for
certain values of $q$ in Table~\ref{tab:xi}. In order to see the first order
nature of the transition in numerical simulations the sizes studied must be
larger than the critical correlation length which suggests that we should take
$ q \ge 16$.

\begin{table}
\begin{center}
\begin{tabular}{|r|l|}
\hline
q   &   $\xi_{\rm crit}$  \\
\hline\hline
4   & $ \infty$ \\
5   &   2512  \\
6   &   158.9  \\
8   &   23.88  \\
12  &    6.548 \\
16  &    3.746 \\
24  &    2.155 \\
32  &    1.608  \\
\hline
\end{tabular}
\end{center}
\caption{
The values of the correlation length at the critical point of the pure system
for certain values of $q$. These results are obtained from expressions in
Ref.~\protect\onlinecite{buffenoir}.
}
\label{tab:xi}
\end{table}

\begin{figure}
\begin{center}
\epsfxsize=\columnwidth
\epsfbox{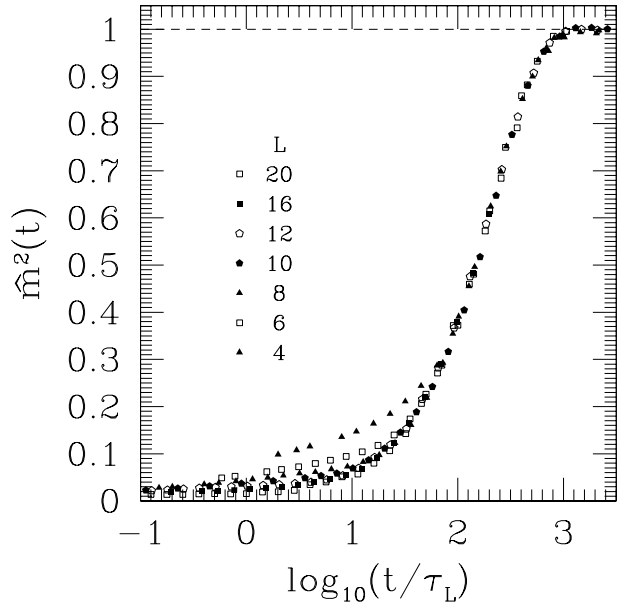} 
\end{center}
\caption{
Results for $\widehat{m}(t)^2$ for the pure system for $q=24$. The horizontal
axis $t/\tau_L$, where $\tau_L$ is determined for each lattice size by
requiring
that the data collapses as well as possible. The resulting values
for the $\tau_L$ are plotted in
Figs.~\ref{pure_24_tau_act} and \ref{pure_24_tau}.
}
\label{pure_24} 
\end{figure}

Data for $q = 24$ for $\widehat{m}(t)^2$
is shown in Fig.~\ref{pure_24}. The data collapse is good
except for very short times, which is presumably not in the scaling regime.
Interestingly, the data for $q=3$
in Fig.~\ref{pure_3_plus_tau}, does scale even at
very short times. 

\begin{figure}
\begin{center}
\epsfxsize=\columnwidth
\epsfbox{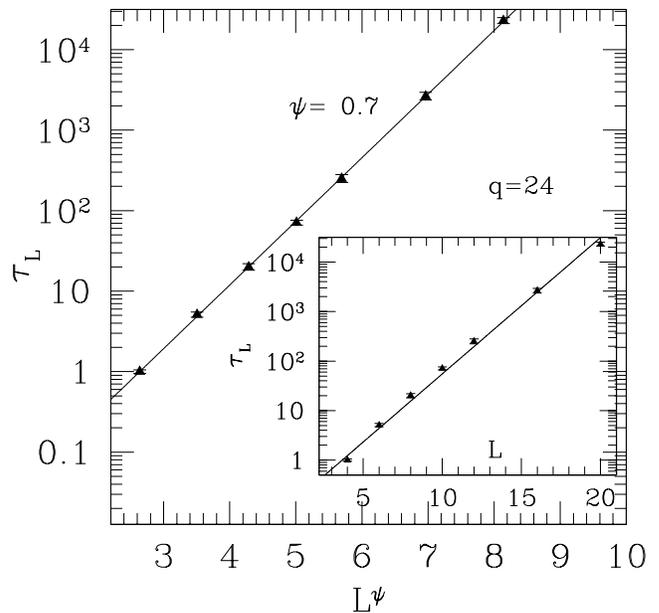} 
\end{center}
\caption{
% Small change
A plot of $\tau_L$ (on a log scale)
against $L^\psi$ for the pure system for $q=24$.
The data fits a straight line well,
indicating activated scaling, though fitting gives $\psi = 0.7 \pm 0.1$,
rather than the asymptotic value of $1$. The inset shows a plot assuming
$\psi=1$. Distinct curvature indicates a less good fit as discussed in
the text.
}
\label{pure_24_tau_act} 
\end{figure}

\begin{figure}
\begin{center}
\epsfxsize=\columnwidth
\epsfbox{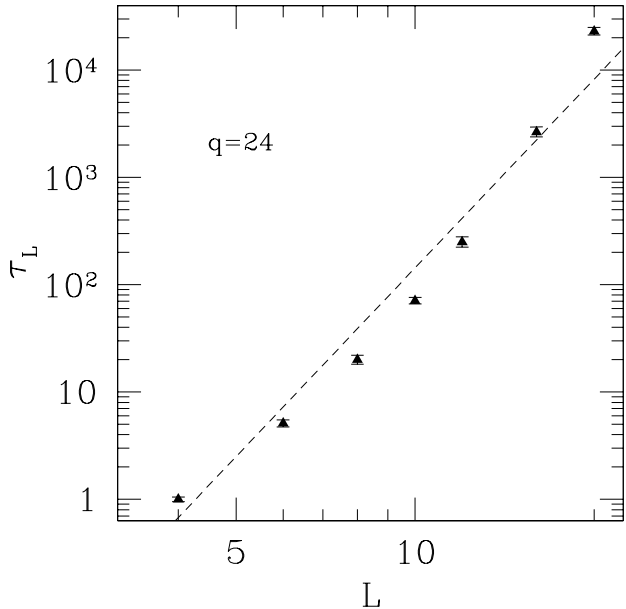} 
\end{center}
\caption{
A  log-log plot of $\tau_L$ against $L$ for the pure system for $q=24$.
The data shows pronounced curvature,
indicating that conventional dynamical scaling does not work.
}
\label{pure_24_tau} 
\end{figure}

Activated and conventional scaling plots for $q=24$ are shown in
Figs.~\ref{pure_24_tau_act}
and \ref{pure_24_tau} respectively. The activated scaling fit works well, but
with a barrier exponent $\psi \simeq 0.7 \pm 0.1$, rather than the expected
value of $1$. The inset shows the data plotted with $\psi = 1$. The fit has a
probability (Q-factor) of $6.1 \times 10 ^{-12}$, which is very low, whereas
the fit with $\psi=0.7$ has a Q-factor of $0.92$, which is good.  Presumably
the value $\psi=0.7$ is only an effective exponent which fits the data for the
range of sizes studied, since one expects to see $\psi=1$ for sufficiently
large sizes.  As shown in Table~\ref{tab:xi}, the correlation length at
criticality decreases with increasing $q$ so one expects to be closer to the
asymptotic value of $\psi$ at large $q$ for the modest range of sizes that we
can simulate.
Our
results are consistent with this since we find effective values of $\psi$
equal to $0.55, 0.7 $ and $0.9$  for $q= 16, 24 $ and $32 $, respectively.
For $q=8$, activated scaling only worked for $\psi$ around 0.1 or
less, which is sufficiently small that it is not significantly different
from conventional scaling.  Thus, although one expects activated scaling
asymptotically for $q=8$, one is far from this regime for the range of sizes
that we can study. This is not surprising since the 
the correlation length at criticality, shown in Table~\ref{tab:xi},
is about 24 lattice spacings. 
Recently, \"Ozo\u{g}uz et al.\cite{ozoguz} have studied the dynamics of the pure
Potts model at criticality for $q=6$ and 7. By using the Wolff\cite{wolff}
algorithm, they are able to study larger sizes than us, and they also
incorporate corrections to finite size scaling. In this way, they find that
their results are consistent with activated scaling with $\psi = 1$. 

The main conclusion from this section is that activated dynamics
is only seen easily in the {\em pure}\/ model for $q \ge 16$. 
To
determine if the {\em random}\/ model has activated dynamics, we expect
that one should choose $q$ to
be {\em larger}\/ than the value where it can be seen for the pure system, i.e.
we need $q > 16$.

\section{Results for the random system}
\label{sec:random}

\subsection{Dynamics of the averaged order parameter}
\label{subsec:average}

\begin{figure}
\begin{center}
\epsfxsize=\columnwidth
\epsfbox{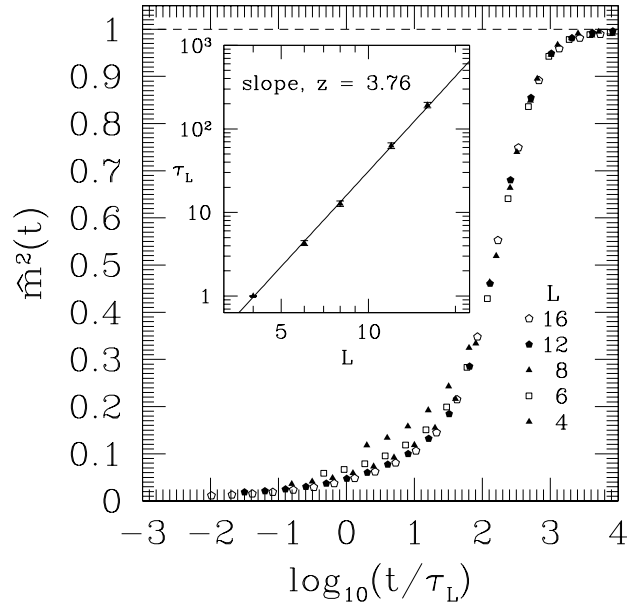} 
\end{center}
\caption{
Results for $\widehat{m}(t)^2$ for the random system with the Olson-Young
distribution, Eq.~(\ref{px}),  for $q=24$. The horizontal axis $t/\tau_L$, where
$\tau_L$ is determined for each lattice size by requiring that the data
collapses as well as possible.
Results for the $\tau_L$ are shown in the inset and
Fig.~\ref{oy_24_tau_act}.
The inset shows a
log-log plot of $\tau_L$ against $L$.
%for the random system with the
%Olson-Young distribution, for $q=24$.
The straight line fit works well and gives a slope
of $z =  3.76 \pm 0.04$.
}
\label{oy_24_plus_tau} 
\end{figure}

Data for $\widehat{m}(t)^2$ for the random system with the Olson-Young
distribution, Eq.~(\ref{px}), for $q=24$ is shown in
Fig.~\ref{oy_24_plus_tau}. The
relaxation times, $\tau_L$ have been determined by requiring that the data for
large times collapses, (with $\tau_4$ arbitrarily set to unity), and are shown
on a log-log plot in the inset to Fig~\ref{oy_24_plus_tau}.
The fit is good and the slope gives
the dynamical exponent, 
\begin{equation}
z =  3.76 \pm 0.04 \qquad (q = 24) .
\label{eq:z_oy_q24}
\end{equation}

%\begin{figure}
%\begin{center}
%\epsfxsize=\columnwidth
%\epsfbox{oy_24_tau.eps} 
%\end{center}
%\caption{
%A  log-log plot of $\tau_L$ against $L$ for the random system with the
%Olson-Young
%distribution, for $q=24$. The straight line fit works well and gives a slope
%of $z =  3.76 \pm 0.04$.
%}
%\label{oy_24_tau} 
%\end{figure}

\begin{figure}
\begin{center}
\epsfxsize=\columnwidth
\epsfbox{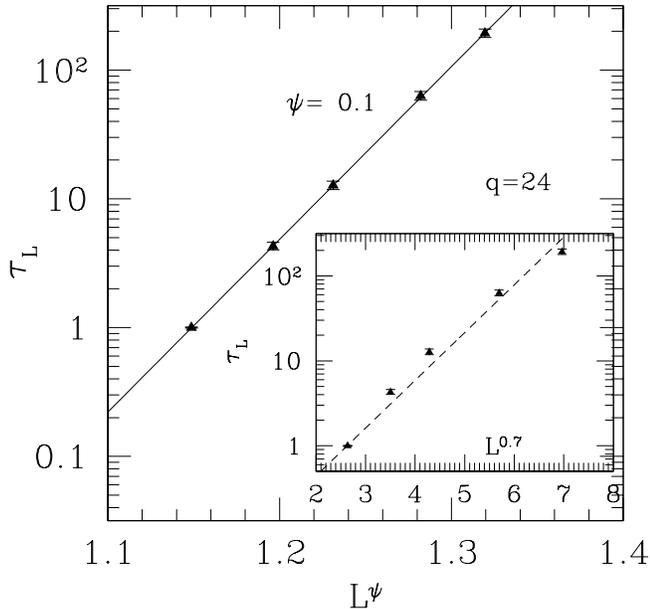} 
\end{center}
\caption{
An activated scaling plot of $\tau_L$ for the random system with the
Olson-Young
distribution, for $q=24$. The only fits which work well have very small
$\psi$, $\psi = 0.1$ is shown here, which, for the range of sizes studied,
is not significantly different from the power law
fit shown in the inset to
Fig.~\ref{oy_24_plus_tau}. The inset shows a plot with $\psi=0.7$,
which is the best value for the pure system, see Fig.~\ref{pure_24_tau_act},
but which clearly doesn't work for the random case
because of the pronounced curvature.
}
\label{oy_24_tau_act} 
\end{figure}

We have also attempted to scale the data using activated dynamical scaling.
However, we find only satisfactory fits are for $\psi$ very small, see
Fig.~\ref{oy_24_tau_act}, which are not significantly different from the power
law fit in the inset to Fig.~\ref{oy_24_plus_tau}.
The inset to Fig.~\ref{oy_24_tau_act} shows a plot with a larger value of
$\psi$, 0.7. The large curvature indicates that this does not work.

It appears, then, that the random Potts model has conventional dynamical
scaling. However, when we discuss distributions of relaxations times in 
Sec.~\ref{subsec:distribution} we shall see that the situation is rather more
complicated, and the true asymptotic behavior is perhaps not clear. 

\begin{figure}
\begin{center}
\epsfxsize=\columnwidth
\epsfbox{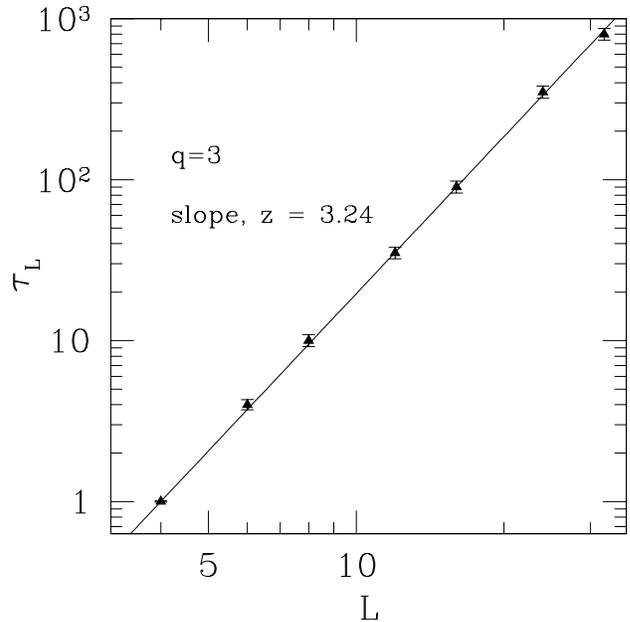} 
\end{center}
\caption{
A  log-log plot of $\tau_L$ against $L$ for the random system with the
Olson-Young
distribution, for $q=3$. The straight line fit works well and gives a slope
of 
$z =  3.24 \pm 0.03$.
}
\label{oy_3_tau} 
\end{figure}

We have also investigated the case of $q=3$, which is unlikely to have
activated dynamical scaling, since the pure system has a continuous
transition, and indeed we find that conventional scaling works well, see
Fig.~\ref{oy_3_tau}. The fit gives 
\begin{equation}
z =  3.24 \pm 0.03 \qquad (q = 3) .
\label{eq:z_oy_q3}
\end{equation}
From Eqs.~(\ref{eq:z_oy_q24}) and (\ref{eq:z_oy_q3}) it appears that
$z$ increases with increasing $q$
and so our results
are compatible with the
value $3.41 \pm 0.06$ obtained by Pan et al.\cite{pan} for $q=8$.

\subsection{Distribution of relaxation times}
\label{subsec:distribution}

Since disorder is relevant for all $q$ greater than two, it is
expected\cite{ah96,ahw98,wd98} that static quantities are not self-averaging
at the critical point. Presumably dynamical quantities like the relaxation
time are also not self-averaging but this does not seem to have been much
discussed up to now. 
We have therefore also studied the \emph{distribution} of relaxation
times for the $q=3$ and $q=24$ Potts models.
For each realization of the disorder (i.e. \emph{sample}), we
generate the time dependent squared magnetization, normalized by
the equilibrium value for
that sample
(which we get with a good precision thanks to the
Wolff algorithm), i.e.
\begin{equation}
\widetilde{m}(t)^2 = { \langle m(t)^2 \rangle \over
\langle m(\infty)^2 \rangle } .
\label{m2samp}
\end{equation}
The thermal average is obtained by repeating the simulation
$10^4$ times with the same random bonds but starting from
different initially disordered
spin configurations.

We find that the {\em shape} of 
$\widetilde{m}(t)^2$ is strongly sample-dependent.
We have tried various definitions to determine a
relaxation time $\tau$ for a single sample:
the time necessary for $\widetilde{m}(t)^2$ to reach a fixed value (0.3,
0.4, 0.5, 0.6, 0.7, 0.8 and 0.9) and the ``integrated time'' defined by
$\int_0^\infty (1-\widetilde{m}^2(t)) d t$.
We checked that all these times,
yield the same scaling behavior.
The best choice, which minimizes the error bars, is the time
to reach 0.60, which will be used from now on.

We have investigated the distribution of relaxation times for $q=3$ and 24.
However, since we need to repeat the dynamical evolution
many times ($10^4$ in practice), and since we need to repeat this for many
samples, the range of sizes that we could study is rather restricted,
$L \le 12$. The number of samples used
for the different sizes and $q$ values is given in table
Table~\ref{tab:numbersamples}.

\begin{table}
\begin{center}
\begin{tabular}{|r|r|l|}
\hline
q  & L  & samples \\
\hline\hline
3  & 4  & 2241 \\
3  & 6  & 1600 \\
3  & 8  & 1537 \\
3  & 12 & 105 \\
\hline
24 & 4  & 2739 \\
24 & 6  & 2223 \\
24 & 8  & 868 \\
24 & 12 & 1012 \\
\hline
\end{tabular}
\end{center}
\caption{
Parameters of the simulations used to study the distribution of relaxation
times for the OY distribution for different values of $q$ and lattice size
$L$.
}
\label{tab:numbersamples}
\end{table}

\begin{figure}
\begin{center}
\epsfxsize=\columnwidth
\epsfbox{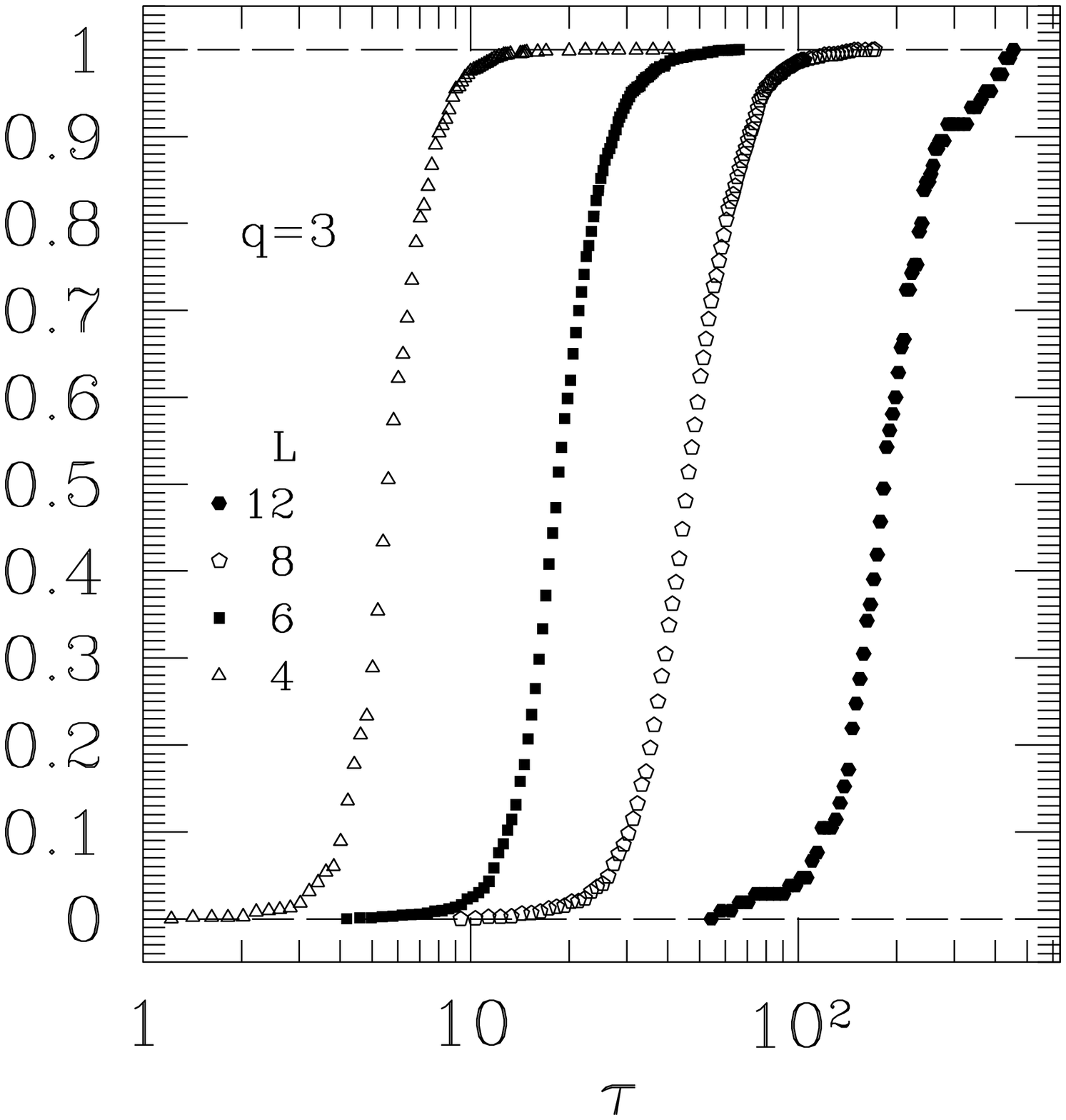}
\end{center}
\caption{
Cumulative distribution functions of $\tau$
for the Olson-Young distribution of random bonds, $q=3$. 
}
\label{cdf_tau_oy_q3}
\end{figure}

\begin{figure}
\begin{center}
\epsfxsize=\columnwidth
\epsfbox{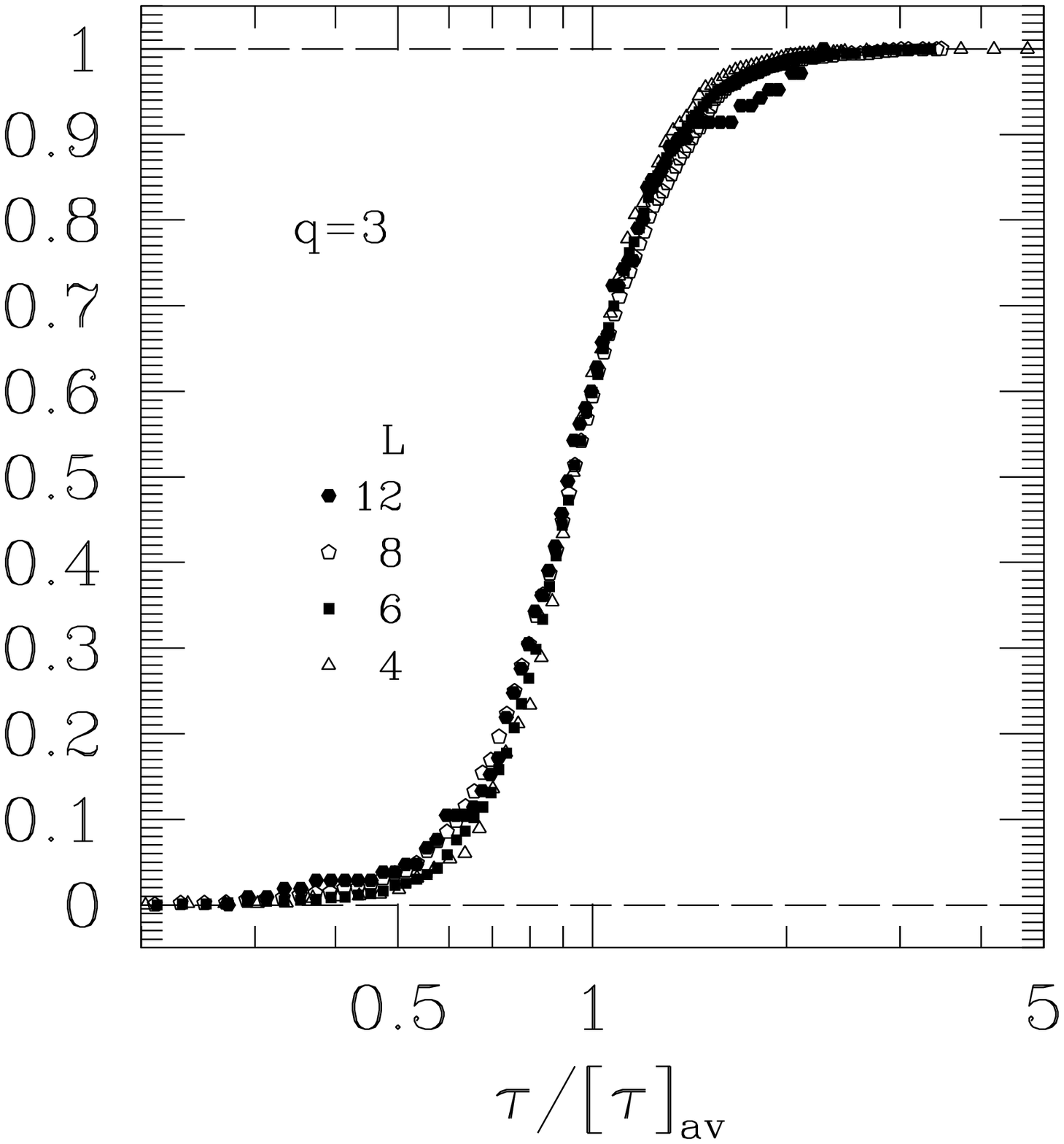}
\end{center}
\caption{Cumulative distribution functions of $\tau/[\tau]_{\rm av}$
for the Olson-Young distribution of random bonds, $q=3$. This rescaling
works
}
\label{cdf_tau_oy_q3_rescale}
\end{figure}

\begin{figure}
\begin{center}
\epsfxsize=\columnwidth
\epsfbox{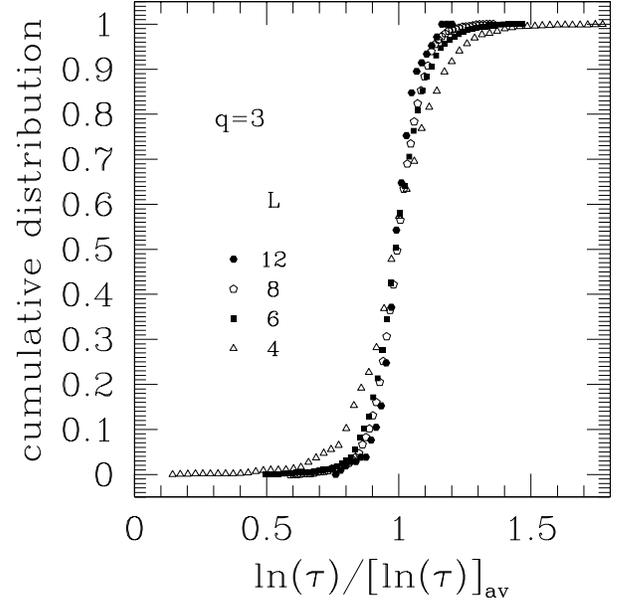}
\end{center}
\caption{Cumulative distribution functions of
$\ln \tau/ [\ln \tau]_{\rm av}$
for the Olson-Young distribution of random bonds, $q=3$. This rescaling
doesn't work since the distribution get narrower for larger $L$.
}
\label{cdf_logtau_oy_q3_rescale}
\end{figure}

\begin{figure}
\begin{center}
\epsfxsize=\columnwidth
\epsfbox{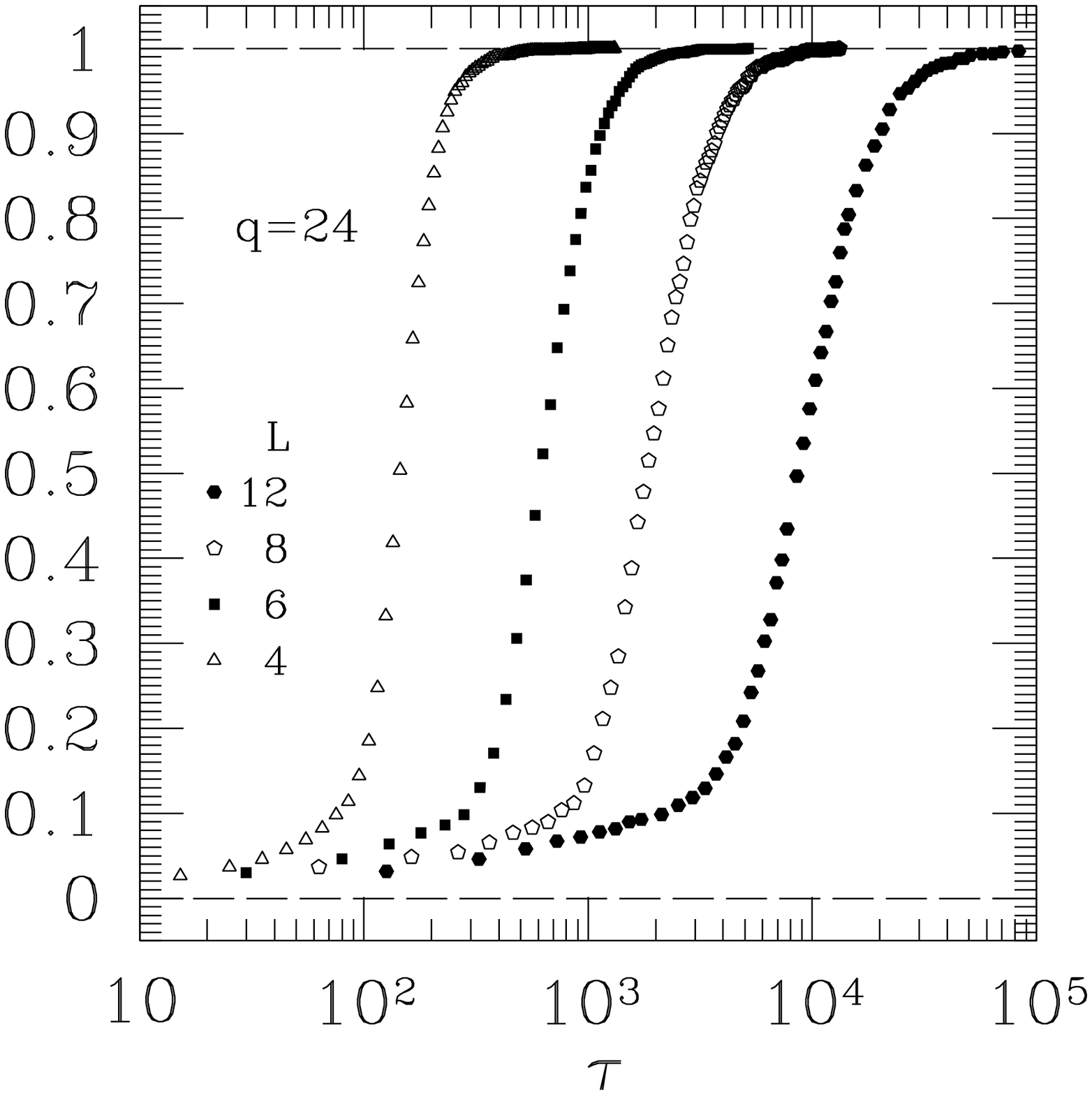}
\end{center}
\caption{
Cumulative distribution functions of $\tau$
for the Olson-Young distribution of random bonds, $q=24$. 
}
\label{cdf_tau_oy_q24}
\end{figure}

\begin{figure}
\begin{center}
\epsfxsize=\columnwidth
\epsfbox{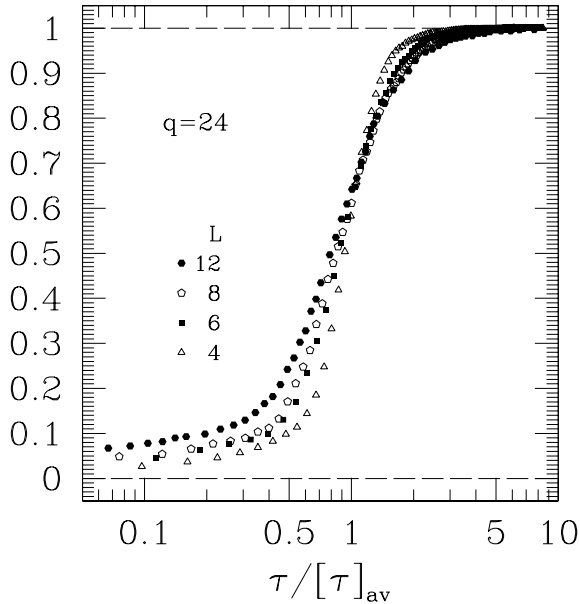}
\end{center}
\caption{Cumulative distribution functions of $\tau/[\tau]_{\rm av}$
for the Olson-Young distribution of random bonds, $q=24$. This rescaling
doesn't work.
}
\label{cdf_tau_oy_q24_rescale}
\end{figure}

\begin{figure}
\begin{center}
\epsfxsize=\columnwidth
\epsfbox{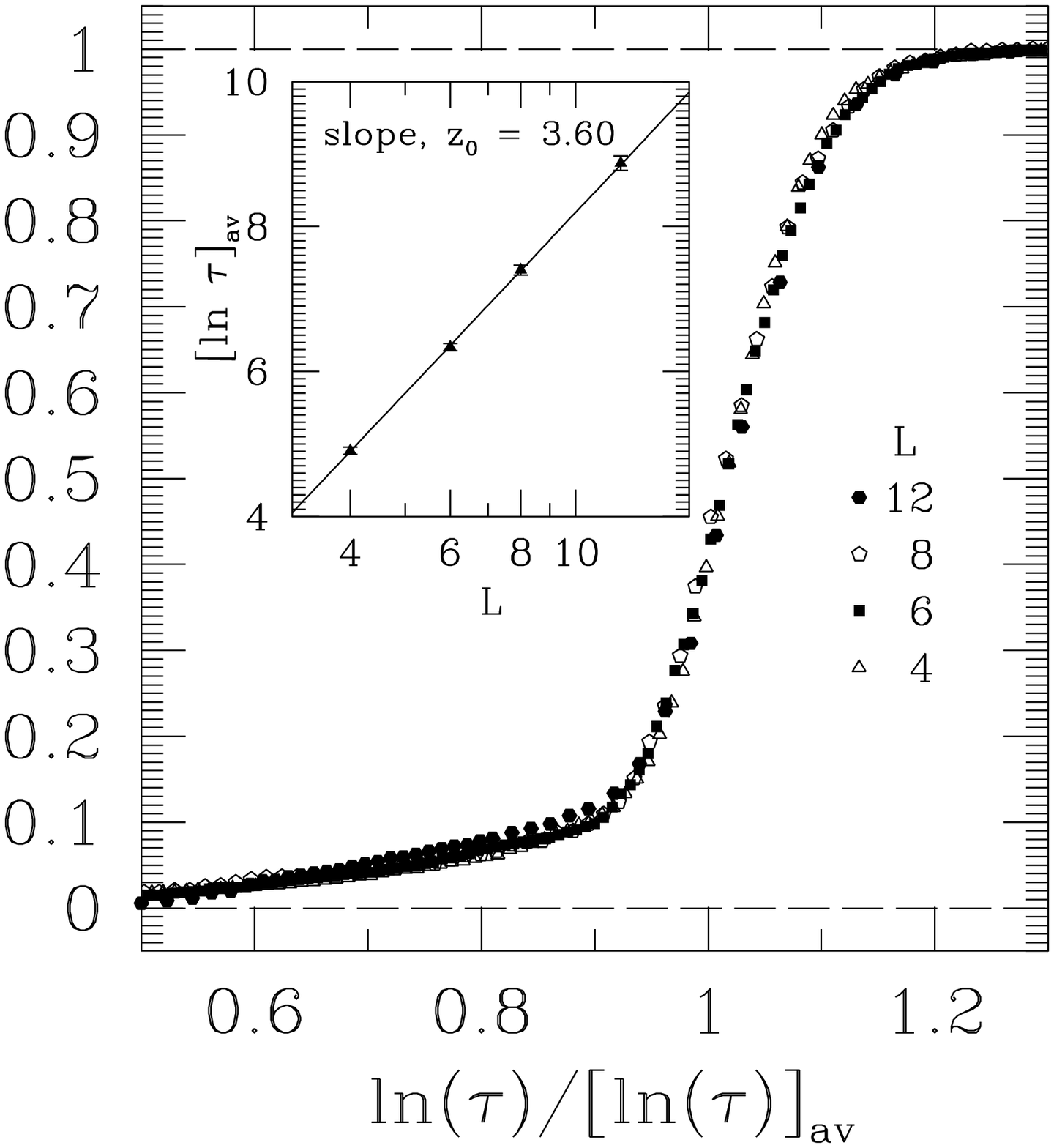}
\end{center}
\caption{Cumulative distribution functions of
$\ln \tau/[\ln \tau]_{\rm av}$
for the Olson-Young distribution of random bonds, $q=24$. This rescaling
works, except for the shortest times.
The inset shows the resulting
$[ \ln \tau ]_{\rm av}$ plotted against $\ln L$, which gives a slope of
$z_0 = 3.60$. 
}
\label{cdf_logtau_oy_q24_rescale_plus_fit}
\end{figure}

Fig.~\ref{cdf_tau_oy_q3}
shows data for the {\em cumulative}\/ distribution of $\tau$ for
$q=3$ for different sizes with a logarithmic horizontal scale.
One sees that the distributions for different sizes are shifted horizontally
but otherwise look very similar. This is
confirmed by Fig.~\ref{cdf_tau_oy_q3_rescale} in which the values for $\tau$
are scaled by the average
relaxation time for each size. The data collapses moderately well,
indicating that $\tau/[\tau]_{\rm av}$ has a (non-trivial)
distribution which is independent
of size, and which is presumably also universal. Hence all powers of moments
of the form $[ \tau^n ]_{\rm av}^{1/n}$, (and the corresponding cumulants)
all vary with $L$ in the same way as the mean, i.e. as $L^z$.
In particular, the relative variance,
\begin{equation}
R_\tau = {[\tau^2]_{\rm av} - [\tau]^2_{\rm av} \over [\tau]^2_{\rm av}  } ,
\label{rtau}
\end{equation}
is a constant for $L \to \infty$ for $q=3$. This is analogous to the lack of
self-averaging in {\em static}\/ quantities at the critical point which has
been discussed before\cite{ah96,ahw98,wd98}.
Note that the
distribution of $\ln \tau / [\ln \tau ]_{\rm av}$
is sharp at large $L$, i.e. $\ln \tau / [\ln \tau ]_{\rm av}$
is self-averaging. We show the sharpening of the distribution of
$\ln \tau / [\ln \tau ]_{\rm av}$ in Fig.~\ref{cdf_logtau_oy_q3_rescale}.

The average relaxation time, $[ \tau]_{\rm av}$, defined in this subsection is
not precisely the same as the relaxation time $\tau_L$ defined in terms of the
dynamics of the square of the
average magnetization in Sec.~\ref{subsec:average}.
However, one expects that they should be proportional to each other, and indeed
this the case since a fit of $\ln [\tau]_{\rm av}$ against $\ln L$
gives a slope of $z = 3.39 \pm 0.10$, essentially in agreement with
Eq.~(\ref{eq:z_oy_q3}).

Fig.~\ref{cdf_tau_oy_q24}
shows data for the cumulative distribution of $\tau$ for
$q=24$. In contrast to the data for $q=3$ in Fig.~\ref{cdf_tau_oy_q3},
the curves for larger sizes are
not only shifted to the right but also become broader. Hence the
data does not collapse in a plot of  $\tau/[\tau]_{\rm av}$ as
shown in Fig.~\ref{cdf_tau_oy_q24_rescale}.

However, if we consider the
distribution of the {\em logarithm}\/ of $\tau$ then a plot
of $\ln \tau / [\ln \tau ]_{\rm av}$ does scale, as shown in
Fig.~\ref{cdf_logtau_oy_q24_rescale_plus_fit}. 
Hence for $q=24$, we find that
$R_\tau$ diverges for $L \to\infty$, whereas $R_{\ln \tau}$ is finite, where 
$R_{\ln \tau}$ is defined in a similar way to $R_\tau$ in Eq.~(\ref{rtau}).
Hence for $q=24$, but not $q=3$, one can say that is is the {\em barriers}\/
which have a finite relative variance. 
Since Fig.~\ref{cdf_logtau_oy_q24_rescale_plus_fit} shows that
$\ln \tau $ is the appropriate scaling
variable for $q = 24$ we need to see how the average of
this quantity varies with size. Hence, in the inset to
Fig.~\ref{cdf_logtau_oy_q24_rescale_plus_fit},
we show a plot of $[\ln \tau ]_{\rm av}$ against $\ln L$, which
works quite well with a slope $3.60 \pm 0.09$.

If we define, possibly different, exponents $z_n$ by
$[\tau^n]^{1/n}_{\rm av} \sim L^{z_n}$
then the slope in the inset to
Fig.~\ref{cdf_logtau_oy_q24_rescale_plus_fit} is $z_0$.
Since $[\ln \tau ]_{\rm av} \sim \ln L$ we have conventional, rather than
activated, dynamics
because in the latter case one would
have $[\ln \tau ]_{\rm av}
\sim L^\psi$. This is also what we found in Sec.~\ref{subsec:average}.
However, since we now see that the scaling variable is
is $\ln \tau$ rather than $\tau$, we need to discuss further the
behavior of 
averages of powers of $\tau$. We shall see that this additional
analysis implies that 
the results in Sec.~\ref{subsec:average} for $q=24$ may not describe the
asymptotic behavior for large $L$.

%\begin{figure}
%\begin{center}
%\epsfxsize=\columnwidth
%\epsfbox{z0_q24.eps}
%\end{center}
%\caption{
%A plot of $[ \ln \tau ]_{\rm av}$ against $\ln L$, which gives a slope of
%$z_0 = 3.60$. 
%}
%\label{z0_q24}
%\end{figure}

Because $\ln \tau$ is the scaling variable, we can write
\begin{eqnarray}
[\tau^n]_{\rm av} & = & [ e^{n \ln \tau} ]_{\rm av} \nonumber \\ 
& = & \int e^{n \ln \tau} f\left( {\ln\tau \over [\ln \tau]_{\rm av}} \right)
{d \ln \tau \over [\ln \tau]_{\rm av} } \nonumber \\
& = & \int e^{ nx [\ln \tau]_{\rm av}} f(x) \, dx  \nonumber  \\
& = & \int e^{ nx (a + z_0 \ln L)} f(x) \, dx , 
\label{lntau_scaling}
\end{eqnarray}
where $f(x)$ is the scaling function for $ x \equiv
\ln\tau / [\ln \tau]_{\rm av}$, and the last equality has used the fit shown
in the inset to Fig.~\ref{cdf_logtau_oy_q24_rescale_plus_fit}.

From Eq.~(\ref{lntau_scaling})
we see that the behavior of $[\tau^n]_{\rm av}$ for large
$L$ depends on the form of the scaling function $f(x)$ for large
$x$. If there is a very sharp cutoff at $x^\star$ say, then the integral is
dominated by values in the vicinity of the cutoff and so $[\tau^n]_{\rm av}
\sim e^{ nx^\star (a + z_0 \ln L)} \sim L^{n x ^\star z_0}$, which
gives $z_n = x^\star
z_0$, independent of $n$, for $n > 0$. This is conventional dynamical scaling,
except that the value of $z_0$ is different from that of $z_n$ with $n > 0$.
However,
more generally, $[\tau^n]_{\rm av}^{1/n}$
would not vary as $L^{z_n}$, and hence would not correspond to
conventional dynamical scaling. For example, a Gaussian form for $f(x)$ would
give $[\tau^n]_{\rm av} \sim \exp( {\rm const.}\ (\ln L)^2 )$, which increases
with $L$ faster than than any power but is slower than exponential. This
behavior is 
in between conventional and activated dynamics.

We have tried to estimate the behavior of $[\tau^n]_{\rm av}$ for large $L$
by noting that
\begin{equation}
[\tau^n]_{\rm av} = g(-i n [\ln \tau]_{\rm av}) ,
\label{ancont}
\end{equation}
where
\begin{equation}
g(k) = \int_0^\infty e^{i k x} f(x) \, dx ,
\end{equation}
is the Fourier transform of the scaling function $f(x)$. Since $f(x)$ is
normalized and has mean unity by definition, we can write
\begin{equation}
g(k) = \exp(i k + u(k)) ,
\label{gk}
\end{equation}
where
\begin{equation}
u(k) = \sum_{n=2}^\infty { \langle x^n \rangle_c (i k)^n \over n!} ,
\end{equation}
in which $\langle x^n \rangle_c$ denotes the {\em cumulant}\ average of
$x^n$. Hence, if we can determine the {\em analytic}\/ form of $u(k)$ for large
$k$, we can analytically continue to the imaginary axis and thereby obtain
$[\tau^n]_{\rm av}$.

Analyzing
the data in 
Fig.~\ref{cdf_tau_oy_q24}, we find that 
the first few cumulants of $f(x)$ are roughly of the form $(-0.13)^n$.
If we use this form then
\begin{equation}
u(k) = \exp(-0.13 i k) - 1 +0.13 i k ,
\end{equation}
which, from Eqs.~(\ref{ancont}) and (\ref{gk}), gives for $L \to \infty$
\begin{equation}
[\tau^n]_{\rm av}
\sim \exp(1.13 n z_0 \ln L)  = L^{1.13 n z_0} ,
\end{equation}
where we have used the fit in the inset to
Fig.~\ref{cdf_logtau_oy_q24_rescale_plus_fit}. This corresponds to the
``sharp cut-off'' mentioned above, i.e. one has dynamical scaling
(except that $z_0$ is different from $z_n$ for $n > 0$).
However, the data is not good enough to be able to make this analytic
continuation with any confidence. 

As a result of this analysis of the distribution of $\ln \tau$, we infer that
the value of $z$ for $q=24$ found above
from $[\tau]_{\rm av}$
(and from $\tau_L$ in Sec.~\ref{subsec:average},
see the inset to Fig.~\ref{oy_24_plus_tau})
{\em may}\/ only represent an
effective exponent, valid for fairly small sizes.

\section{Conclusions}
\label{sec:conclusions}
We have studied the dynamics of the $q$-state random bond Potts ferromagnet at
the critical point in two-dimensions for $q=3$ and $q=24$. In both cases we
find conventional dynamical scaling, and our results for the dynamical
exponent are compatible with those of Ref.~\onlinecite{pan}.
However, for $q=3$, we find that the
reduced variance of the relaxation time $\tau$ tends to a finite value for
large $L$, while for $q=24$ it is the reduced variance of $\ln \tau$ which tends
to a finite value. Since $\ln \tau$ is proportional to a barrier height, our
results for $q=24$ imply that there is scaling for
the distribution of barrier
heights but the characteristic barrier height scales with the log of
the system size.
%Equivalently, for $q=24$ one can say that the reduced variance of the
%barrier heights is finite. 

We do not have an intuitive explanation for this
difference in behavior between the two cases, but it may be related to the
transition of the pure system being first order for $q=24$ and second order
for $q=3$. It is currently unclear whether the 
scaling of the distribution of barrier heights for $q=24$
implies that dynamical scaling is actually activated for $ L \to \infty$, in
which case our finite value of $z$ would just be an effective exponent valid
for rather small sizes.

\acknowledgments
We acknowledge support from the NSF through grant DMR 0086287. We would like
to thank Terry Olson for helpful discussions and for making available his code
for the Wolff algorithm.
One of us (APY) would also like to
thank John Cardy and David Huse for helpful discussions and comments on an
early version of the manuscript.  The
other (CD) would like to thank UCSC for hospitality while part of this work
was performed.


\begin{thebibliography}{99}
\bibitem{correction}
    In the published version, Ref.~\onlinecite{terry}, the space between the
    factor $2/\pi$ and the rest of the expression in Eq.~(\ref{px}) is
    mysteriously missing, though the space was present in the version
    submitted to the journal, see cond-mat/9903068. According to this typo, 
    $\pi$ would only multiply the factor of $(1-x)^2$ in the
    denominator, which is incorrect.
\end{thebibliography}
\end{document}